\documentclass[aps,pra,twocolumn,showpacs]{revtex4}
\usepackage{epsfig}
\usepackage{epsf}
\usepackage{bm}
\usepackage{mathrsfs}
\usepackage{amsmath}
\usepackage{amssymb}
\usepackage{graphicx}
\usepackage{amsfonts}
\usepackage{amsthm}
\usepackage{color}
\usepackage{dcolumn}
\usepackage{txfonts}

\begin{document}
\title{Non-Markovian dynamics and strong coupling between atomic transitions and a waveguide continuum edge}

\author{Ting Chen}
\author{Ren-Bao\ Liu}
\email{rbliu@phy.cuhk.edu.hk}
\affiliation{Department of Physics and Centre for Quantum Coherence, The Chinese University of Hong
Kong, Shatin, New Territories, Hong Kong, China}
\date{\today}

\begin{abstract}
In quantum communication and distributed quantum computing, one-dimensional waveguides provide directional
transfer of quantum information. A single-mode waveguide has a density-of-states singularity
at the lower cut-off frequency, which resembles sharp resonances of a cavity but with non-Markovian dynamics.
Thus we put forward schemes of coupling atomic transitions and a waveguide continuum edge.
We first present a scheme of spin-photon quantum interface operating in the non-Markovian regime for a $\Lambda$-type three-level system coupled to a waveguide.
Then we show that strong coupling between atomic transitions and a waveguide continuum edge
can lead to vacuum Rabi oscillations and bound polariton states.
\end{abstract}

% 03.67.Lx Quantum computation architectures and implementations;
% 42.50.ct quantum description of interaction of light and matter;
% 42.79.Gn optical waveguides and couplers

%\pacs{81.40.Pq; 46.55.+d; 62.20.Qp}
% 46.55.+d Tribology and mechanical contacts
% 62.20.Qp Friction, tribology and hardness in mechanical properties of solids
% 81.40.Pq Friction, lubrication and wear in materials science

\maketitle

%==================================================================
\section{Introduction}
\label{intro}

Quantum networks are essential in quantum communication and distributed quantum
computing~\cite{David,PhysRevLett783221,PhysRevA.67.032305,liu2010quantum}.
A quantum network consists of local nodes and connecting
quantum channels. The stationary qubits in local nodes can be
provided by
trapped ions~\cite{ion,PhysRevLett.75.4714,cirac2000scalable}, collective atomic excitations ~\cite{optfleischhauer,PhysRevLett.84.4232},
 quantum dots~\cite{PhysRevLett.83.4204,liu2010quantum,wy}, or solid-state impurities~\cite{PhysRevLett.97.247401,togan2010quantum,buckley2010spin,PhysRevLett.105.177403,PhysRevLett.105.220501}. Single
photons produced by manipulation of atomic transitions in such systems
can form the flying qubits. The quantum information
carried by the flying qubits can be conducted between local nodes through waveguides.
To improve the efficiency of conversion between stationary and flying qubits,
cavities have been adopted to enhance the photon emission~\cite{PhysRevLett783221,wy,PhysRevA.67.032305,liu2010quantum,hennessy2006quantum}.

In this paper, we present schemes of strong coupling between
atomic transitions and waveguide continua without an intermediate
cavity. The basic idea is: at the lower cut-off frequency of
a single-mode waveguide, there is a singularity in the density-of-states
(DOS) of the one-dimensional continuum.
Such a DOS singularity resembles the sharp resonances associated with
discrete states of cavities. The line-shape of the singularity, in contrast to
those of cavity resonances, is highly non-Lorentzian, and therefore the
photon emission at the continuum edge is a non-Markovian process.
Such non-Markovian nature induces interesting memory effects on, e.g.,
spin-photon quantum interfacing. We derive an exact solution for the quantum
interfacing scheme in the non-Markovian regime, extending the control scheme of spin-photon interfaces that works in the Markovian regime~\cite{wy}.
Also, the coupling between atomic transitions and waveguide continuum edges
can lead to vacuum Rabi oscillations and bound polariton states, similar
to atom-cavity quantum electrodynamics in the strong coupling regime.
These results are relevant to recent investigation on coherent coupling between atomic transitions and one-dimensional continuum of photons or plasmonics~\cite{lukin,zhang2009ultra,claudon2010highly}.

There have been a number of proposals~\cite{PhysRevLett783221,optfleischhauer,PhysRevLett.84.4232,PhysRevA.67.032305,wy} and experimental implementations ~\cite{blinov2004observation,PhysRevLett.92.213601,PhysRevLett.93.263601,chou2005measurement}
of quantum interface between stationary qubits and flying photon qubits.
A milestone is the proposal of the state transfer between two atoms in cavities~\cite{PhysRevLett783221}.
The scheme is based on the idea of using two mutually time-symmetric
laser pulses to control the cavity-assisted Raman processes in the two atoms
at the sending and receiving nodes. Such a scheme, however, requires the two
nodes be identical and is sensitive to randomness in atom-cavity couplings.
In order to overcome such obstacles, adiabatic schemes similar to the stimulated
Raman adiabatic passage~\cite{RevModPhys.70.1003} have been proposed
to implement the cavity-assisted Raman processes~\cite{optfleischhauer,PhysRevA.67.032305,PhysRevLett.84.4232}. The adiabatic scheme has the
advantage of robustness against parameter uncertainty but the disadvantage of
slow operation rates. An exact solution of the control pulse for spin-photon interfacing was later
discovered~\cite{wy}, in which
the photon wavepacket can be arbitrarily tailored by shaping the laser pulse that controls the
Raman process in a $\Lambda$-type atomic system coupled to the cavity. By combining the
sending and receiving functions of two spin-photon interfaces, deterministic state transfer and entanglement
between two remote stationary qubits are possible, without requiring two identical nodes. Such a control scheme is particularly suitable for solid-state quantum networks,
where local nodes can hardly have identical parameters.
In the above spin-photon interface schemes, the cavity modes are coupled to
a wide band of photon continuum in free space or waveguides. Thus the
photon sending or receiving can be well described by the input-output theory~\cite{PhysRevA.30.1386}
with Markovian approximation (i.e., Wigner-Weisskopf approximation)~\cite{quantumoptics}.

The quantum interface scheme proposed in this paper uses direct coupling between
atomic transitions and the DOS singularity of a waveguide continuum (without a cavity as
the intermediate). The photon emission/absorption becomes a non-Markovian process
due to the non-Lorentzian lineshape of the photon continuum DOS. We extend the exact solution
in Ref.~\cite{wy} to the general case of non-Markovian coupling.
The exact solution reduces to the input-output theory in the limit of Markovian approximation.
In order to investigate the validity of the
Markovian approximation, we numerically simulate the control scheme for various
operations for different coupling parameters (such as detuning frequencies, coupling constants,
and pulse durations). The quantum interfaces including waveguides are widely used in solid-state systems,
such as cooper pair boxes coupled to a transmission line~\cite{PhysRevLett.95.213001} and single emitters coupled to the surface-plasmon modes of a metal nanowire~\cite{lukin}. Quantum dots can also couple to etched waveguides or 1D
 photonic crystals~\cite{PhysRevLett.87.253902,zhang2009ultra} to form quantum interfaces.
 Our scheme works for those quantum interfaces in the strong coupling regime, where the Markovian approximation may not be valid.

The DOS singularity of a waveguide continuum resembles a sharp cavity resonance. Therefore we envisage
strong coupling between atomic transitions and the continuum edge.
We study the photon emission dynamics of a two-level system, and find vacuum Rabi oscillations of the non-Markovian emission
and a bound polariton state which results in photon localization and incomplete emission. These results are related to
the photon localization phenomena previously discovered in photonic band-gap materials~\cite{PhysRevLett.74.3419,PhysRevLett.95.213001,PhysRevLett.101.100501}.

This paper is organized as follows: In Sec.~\ref{statetransfer}, we explain
the basic idea of quantum interface in the non-Markovian coupling regime,
present an exact solution of the control, and numerically investigate the validity of the
Markovian approximation and the effects of photon leakage.  In Sec.~\ref{photontrap}, we study the spontaneous emission of an exited
two-level system coupled to a one-dimensional waveguide. Sec.~\ref{conclusion} summarizes and concludes the
paper.

%==================================================================
\section{Spin-photon interface by atom-waveguide quantum electrodynamics}
\label{statetransfer}

\subsection{Description of the control scheme}
\label{model}
\begin{figure}[t]
\includegraphics[width=\linewidth]{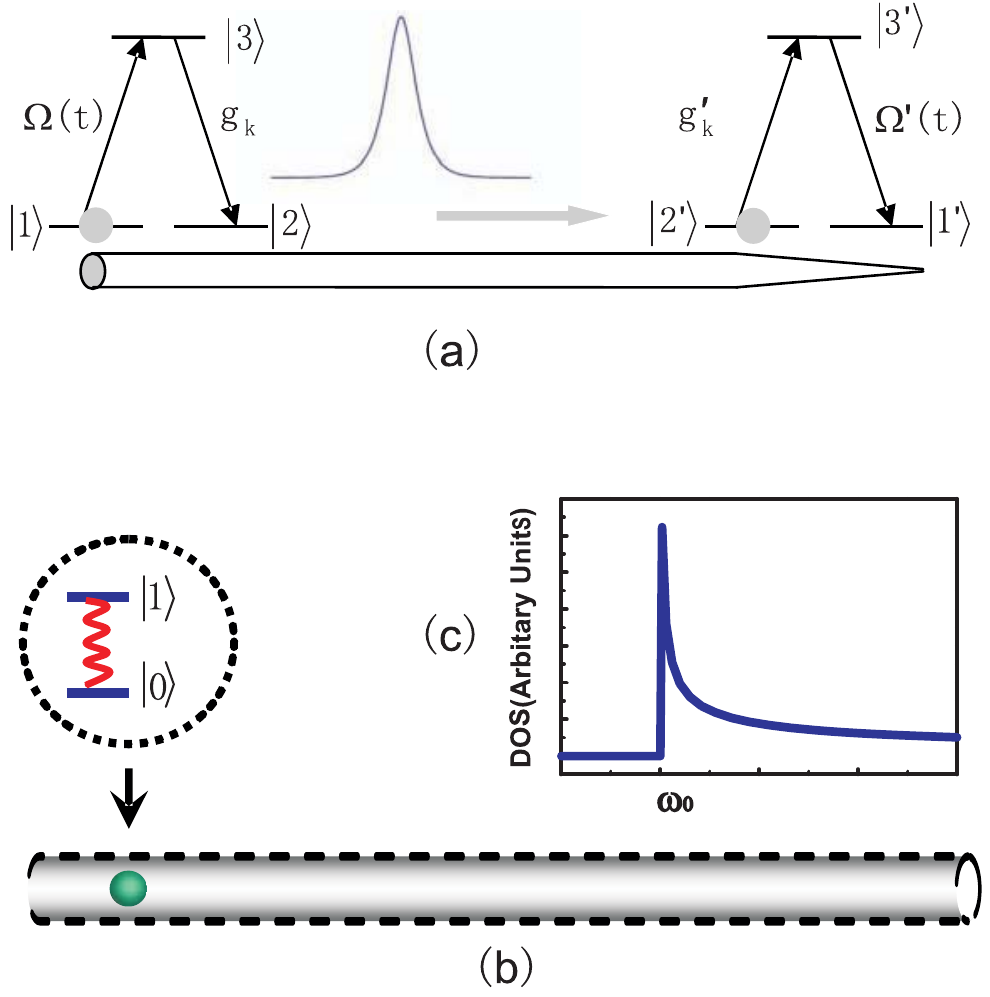}
\caption{Illustration of atom-waveguide quantum electrodynamics.
(a) The local nodes composed of three-level systems.
$|1\rangle$ ($|1^\prime\rangle $) and $|2\rangle$ ($|2^\prime\rangle$)
are two-near degenerate ground states forming the qubit
and $|3\rangle$ ($|3^\prime\rangle$) is an excited state. The state $|2\rangle$ ($|2^\prime\rangle$)
is coupled to the intermediate state $|3\rangle$ ($|3^\prime\rangle$) by the waveguide
modes $\{|k\rangle\}$ with strength $g_k$ ($g^\prime_k$) and the state $|1\rangle$ ($|1^\prime\rangle $) is coupled to
the state $|3\rangle$ ($|3^\prime\rangle$) by a classical pulse with Rabi frequency
$\Omega(t)$ $\left[\Omega^\prime(t)\right]$. These two nodes are connected by the waveguide. The output of the sending node is directed to the receiving
node as its input. (b) The two-level system
embedded in a one-dimensional waveguide. The transition couples to the waveguide modes with
strength $g_k$. (c) Schematic density of states in a
one-dimensional waveguide near the lower cut-off frequency ($\omega_0$).} \label{fig:model}
\end{figure}

Figure~\ref{fig:model}~(a) shows the scheme of quantum interface and state transfer.
The basic idea is to control the emission of a single photon in
the sending node into the directional waveguide and
the absorption of the photon at the receiving node so that quantum information
is transferred between the two stationary qubits~\cite{PhysRevLett783221}.
The stationary qubit in a node is encoded in the two near degenerate ground states
of a $\Lambda$-type three-level system ($|1\rangle$ and $|2\rangle$ in the sending node, and
$|1'\rangle$ and $|2'\rangle$ in the receiving node). The two three-level systems are
connected by a quantum channel formed by a semi-infinite waveguide.
The coupling to the photon continuum in the waveguide
is  mediated by the excited states $|3\rangle$ and $|3'\rangle$ in the sending and receiving nodes, respectively. Note that the two nodes are not required to be identical.
 The driving laser
with Rabi frequency $\Omega$ is resonant to the transition
$|1\rangle\leftrightarrow|3\rangle$.
If the sending node is initially in the state $|1\rangle$, the laser pulse
is designed such that the system is excited to the intermediate state $|3\rangle$
and then relaxes to the state $|2\rangle$  with a photon wavepacket
emitted to the waveguide. The photon wavepacket can be shaped on demand by
designing the control laser pulse $\Omega(t)$.
When the photon arrives at the receiving node, a control laser pulse can be designed to
absorb the photon without reflection (quantum impedance match). The design of the
control pulse at the receiving node for a certain arriving photon wavepacket
can be obtained by time-reversal of a control laser pulse for sending the
time-reversed photon wavepacket~\cite{wy}. The DOS singularity of the photon continuum in
the 1D waveguide provides strong coupling to facilitate the quantum interface (so that the
leakage of quantum information into free space is negligible), which otherwise requires
a cavity to enhance the coupling. Near the continuum edge of
the 1D waveguide, the photon emission and absorption are in general non-Markovian.
A key issue to be addressed in this paper is how to design the control laser pulse $\Omega(t)$
to realize the emission or absorption of an arbitrary photon wavepacket in the
non-Markovian regime, which is the inverse problem of quantum evolution of the coupled atom-waveguide system under time-dependent control.

%\begin{figure}[t]
%\includegraphics[scale=0.45] {waveg.eps}
%\caption{The illustration of the model. A two-level system is
%embedded in a one-dimensional waveguide. (a)State $|1\rangle$ is
%coupled to the intermediate $|0\rangle$ by the waveguide modes with
%strength $g_k$. (b) Schematic diagram of the density of the state in
%one-dimensional waveguide.}\label{fig:waveg}
%\end{figure}

\subsection {Exact solution for control design}
\label{design}

In this subsection, we present the exact solution of the control laser
pulse for an arbitrary photon wavepacket. Here we neglect for the moment
photon leakage into free space through the excited states $|3\rangle$  and $|3^\prime\rangle$.
The control errors induced by the free-space emission will be investigated
later in Sec.~\ref{error}.

The quantum interface is shown in Fig.~\ref{fig:model}(a).
The state $|1\rangle$ is coupled to $|3\rangle$ with $x$ -polarization light. And
the state $|2\rangle$ is coupled to $|3\rangle$ with $y$ -polarization waveguide modes. The
Hamiltonian describing the interaction between the three-level systems
and the waveguide continuum, with rotating-wave approximation and in the
rotating reference frame, is
\begin{align}
H=&\left[\sum_kg_k|3\rangle\langle2|a_ke^{i(\varepsilon_{32}-\omega_k)t}+\Omega(t)|3\rangle\langle1|\right]+\text{h.c.}\nonumber\\
+&\left[\sum_kg^{\prime}_k|3^\prime\rangle\langle2^\prime|a_ke^{i(\varepsilon^\prime_{32}-\omega_k)t}+\Omega^\prime(t)|3^\prime\rangle\langle1^\prime|\right]+\text{h.c.},
\end{align}
%\begin{align}
%H&=\sum_k[g_k|3\rangle\langle2|a_ke^{i(\varepsilon_{32}-\omega_k)t}+g^*_k|2\rangle\langle3|a^\dag_ke^{-i(\varepsilon_{32}-\omega_k)t}]\nonumber\\
%&+\Omega(t)|3\rangle\langle1|+\Omega^*(t)|1\rangle\langle3|+\nonumber\\
%&\sum_k[g^{\prime}_k|3^\prime\rangle\langle2^\prime|a_ke^{i(\varepsilon_{32}-\omega_k)t}+g^{\prime
%*}_k|2^\prime\rangle\langle3^\prime|a^\dag_k
%e^{-i(\varepsilon_{32}-\omega_k)t}]\nonumber\\
%&+\Omega^\prime(t)|3^\prime\rangle\langle1^\prime|+\Omega^{\prime*}(t)|1^\prime\rangle\langle3^\prime|
%\end{align}
where $a_k$ is the annihilation operator of the waveguide mode $|k\rangle$,
the transition
$|2\rangle\leftrightarrow|3\rangle$
 is coupled to the waveguide mode  $|k\rangle$ with strength $g_k$,
the transition $|1\rangle\leftrightarrow|3\rangle$
is driven by a laser pulse with time-dependent Rabi frequency
$\Omega(t)$ with central frequency equal to the transition frequency between
$|3\rangle$ and $|1\rangle$, $\varepsilon_{32}$ is the energy splitting between $|3\rangle$ and $|2\rangle$.
$\omega_k=\sqrt{\omega^2_0+(c/n)^2 k^2}$ is the photon frequency with $\omega_0$ denoting the lower cut-off frequency,
 and $n$ denoting the refractive index of the waveguide. Near the cut-off frequency, the photon dispersion is close to that of a massive particle and the $\text{DOS}=\sqrt{\frac{\omega_0}{2}}\frac{1}{(c/n)}\frac{1}{\sqrt{\omega-\omega_0}}$  has a singularity $[$Fig.~\ref{fig:model}(c)$]$.
Symbols with prime marks have similar meaning but for the receiving node.

The coupled system has an invariant Hilbert subspace expanded by
the basis states $\{|1,2^\prime\rangle
|\text{vac}\rangle,|3,2^\prime\rangle|\text{vac}\rangle,$
$|2,2^\prime\rangle|k\rangle,|2,3^\prime\rangle|\text{vac}\rangle,|2,1^\prime\rangle|\text{vac}\rangle\}$.
Here the front vectors denote the states of the three-level
systems, $|\text{vac}\rangle$ denotes the vacuum state of the waveguide, and $|k\rangle$ stands for the one-photon Fock state of the
waveguide mode of wavevector $k$. The state-transfer works for arbitrary initial state of the sending node,
while the receiving node should be initialized to the state $|2'\rangle$. The state of the system can be written as
\begin{equation}
|\Psi(t)\rangle=\alpha|2,2'\rangle +\beta|\psi(t)\rangle,
\end{equation}
with
\begin{align}
|\psi(t)\rangle=&C_1(t)|1,2^\prime\rangle+C_3(t)|3,2^\prime\rangle+\int_0^{\infty}
C_2(k)|2,2^\prime\rangle|k\rangle \text{d}k\nonumber\\
&+D_3(t)|2,3^\prime\rangle+D_1(t)|2,1^\prime\rangle.
\end{align}
The coefficients satisfy the Schr$\ddot{\text{o}}$dinger
equation
\begin{subequations}
\begin{align}
&\partial_t{C_1(t)}=-i\Omega^*(t)C_3(t), \\
&\partial_t{C_3(t)}=-i\Omega(t)C_1(t)-i\int^{+\infty}_0g_kC_2(k)e^{i\varepsilon_{32}t-i\omega_kt} \text{d}k ,\\
&\partial_t{C_2(k,t)}=-ig^*_k
C_3(t)e^{i\omega_k t-i\varepsilon_{32}t}-ig^{\prime*}_k
D_3(t)e^{i\omega_kt-i\varepsilon^\prime_{32}t},\\
&\partial_t{D_3(t)}=-i\Omega^\prime(t)D_1(t)-i\int^{+\infty}_0g^{\prime}_k C_2(k)e^{i\varepsilon^\prime_{32}t-i\omega_kt} \text{d}k,\\
&\partial_t{D_1(t)}=-i\Omega^{\prime*}(t)D_3(t).
\end{align}
\end{subequations}

If the waveguide is long, we can separate the state transfer into two independent steps, namely, sending and receiving of the photon wavepacket. The emitted photon at the remote future for the sending node can be treated as the incoming photon at the remote past for the receiving node. The output wavepacket of the first node constitutes the input one for
the second node with a time delay. We regard the time
delay $t_0$ as $\frac{L}{v_g}$, where $L$ is the length of the waveguide and $v_g$ is the group
velocity of the photon wavepacket in the
waveguide. The waveguide is long enough so that the second atom is
decoupled from the waveguide during the photon emission in the
sending node. When the wavepacket propagates down the waveguide and
enters the second node, the sending node does not interact with the
wavepacket any more. Therefore, the dynamic equations are decoupled into two parts for the
sending and receiving process respectively,
\begin{subequations}
\begin{align}
\partial_t{C_1(t)}=&-i\Omega^*(t)C_3(t),\\
\partial_t{C_3(t)}=&-i\Omega(t)C_1(t)-i\int^{+\infty}_0g_kC_2(k)e^{i\varepsilon_{32}t-i\omega_kt} \text{d}k,\\
\partial_t{C_2(k,t)}=&-ig^*_k
C_3(t)e^{i\omega_kt-i\varepsilon_{32}t}.\label{eqsendc}
\end{align}
\label{eqsend}
\end{subequations}
for $t\approx 0$ , and
\begin{subequations}
\begin{align}
\partial_t{D_1(t)}=&-i\Omega^{\prime*}(t)D_3(t),\\
\partial_t{D_3(t)}=&-i\Omega^\prime(t)D_1(t)-i\int^{+\infty}_0g^{\prime}_kC_2(k)e^{i\varepsilon^\prime_{32}t-i\omega_kt} \text{d}k,\\
\partial_t{C_2(k,t)}=&-ig^{\prime*}_k D_3(t)e^{i\omega_kt-i\varepsilon^\prime_{32}t},
\end{align}
\label{eqreceiv}
\end{subequations}
for $t\approx t_0$.

For the sending node, we look for a solution of the driving
pule $\Omega(t)$ for an arbitrary output $F(k)$, which determines the
photon wavepacket shape. The boundary conditions are taken as
$C_2(k,-\infty)= 0$ and $C_2(k,+\infty)=F(k)$. This means that the sending node functions under the condition of no incoming photon and generates an outgoing photon wavepacket of the desired shape.
From Eq.~(\ref{eqsendc}), we have
\begin{align}
F(k)=-i\int^{+\infty}_{-\infty}C_3(t) {g_k^*} e^{i\omega_k
t-i\varepsilon_{32} t}\text{d}t.
\end{align}
By inverse Fourier transform, the coefficient
$C_3(t)$ is obtained as
\begin{align}
C_3(t)&=i\int^{+\infty}_{0}\frac{1}{g_k^*}F(k)e^{-i\omega_k
t+i\varepsilon_{32} t}\frac{\text{d}\omega_k}{2\pi},
\end{align}
from which $C_2(k,t)$ is derived from Eq.~(\ref{eqsendc}). To get the coefficient $C_1(t)$, we write it as
$|C_1(t)|e^{i\phi_1(t)}$, where the phase $\phi_1$ is a real number. The
amplitude of $C_1(t)$ is given by the normalization condition
$|C_1(t)|^2=1-|C_3(t)|^2-\int^{+\infty}_0|C_2(k,t)|^2 \text{d}k$ and
the phase

\begin{align}
\phi_1(t)=\int^t_{-\infty}dt^\prime\frac{|C_3(t^\prime)|^2\partial_{t^\prime}\phi_3-\int^{\infty}_0|C_2(k,t^\prime)|^2\partial_{t^\prime}\phi_2
\text{d}k}{|C_1(t^\prime)|^2}.
\end{align}
Finally, from Eq.~(\ref{eqsend}), we derive the driving field
$\Omega(t)$ in terms of all coefficients,
\begin{align}
\Omega(t)=\left[iC^{-1}_3(t)\partial_tC_1(t)\right]^*.
\label{eqomega}
\end{align}

%=====================================================================================
The control of the receiving node is similar to that of the sending node but with different boundary conditions.  The photon emitted by the sending node is fully absorbed by the receiving node. This means the boundary conditions $C_2(k,t_0+t')=F(k)$ for $t'\rightarrow-\infty$ and $C_2(k,t_0+t')=0$ for $t'\rightarrow+\infty$ (here for convenience we have written $t\equiv t_0+t'$). The coupling strength between the atom transitions and the
quantum channel depends on the position of the node through the relation
~\cite{pzoller} $g^{\prime}_k=g_k e^{ikL}$. The coefficient $D_3(t)$
and $C_2(t)$ is expressed as
\begin{subequations}
\begin{align}
&D_3(t^\prime+t_0)=-i\int^{+\infty}_{0}\frac{1}{g^{\prime*}_k}F(k)e^{-i(\omega_k-\varepsilon^\prime_{32})(t^\prime+t_0)}\frac{\text{d}k}{2\pi}\partial_{k}\omega_k ,\\
&C_2(k,t^\prime+t_0)=F(k)-\nonumber\\
&\int^{t^\prime}_{-\infty}dt^{\prime\prime}\int^{\infty}_0
e^{-i(k-k^\prime)L+i(\omega_k-\omega_{k^\prime})(t_0+t^{\prime\prime})}F(k^\prime)\partial_{k^\prime}
\omega_{k^\prime}
\frac{\text{d}k^\prime}{2\pi}.
\end{align}
\end{subequations}
For the receiving process, the control
pulse is obtained as
\begin{align}
\Omega^\prime(t^\prime+t_0)=\left[iD^{-1}_3(t^\prime+t_0)\partial_{t^\prime}D_1(t^\prime+t_0)\right]^*.
\end{align}
By combing the sending and receiving processes, the quantum state transfer can be implemented  and remote entanglement can also be generated~\cite{wy}.
Thus the exact solutions give a general control scheme for quantum interface.

In the Markovian approximation, the equations of motion are decoupled
into the following equations~\cite{PhysRevA.30.1386}
\begin{subequations}
\begin{align}
\partial_t{C_1(t)}&=-i\Omega^*(t)C_3(t),\\
\partial_t{C_3(t)}&=-i\Omega(t)C_1(t)-\frac{\gamma}{2}C_3(t),
\end{align}
\label{eqmak}
\end{subequations}
where the decay rate $\gamma=2\pi
|g_k|^2({\frac{\text{d}k}{\text{d}\omega_k}})$ is a constant,
in which the DOS ($\text{d}k/\text{d}\omega_k$) is taken as the value at the central
frequency of the output wavepacket. The approximation relies on the assumption that the DOS of the
waveguide is flat within the outgoing wavepacket spectral width. In the Markovian approximation, the control pulse $\Omega(t)$ is
still given by Eq.~(\ref{eqomega}), but the coefficients ($C_1$ and $C_3$) are determined by Eq.~(\ref{eqmak}).
%===================================================================
\subsection{Numerical investigation}
\label{numerical}

In this subsection, we numerically study the control design of quantum interface.  In particular, by numerical comparison of the control designs with and without the Markovian approximation, we investigate the validity range of the approximation. The results show that the Markovian approximation is well justified only when the spectral width of the photon wavepacket is much less than the separation of the wavepacket center frequency from the waveguide cut-off frequency. For the moment, the photon leakage
into the free space is assumed to be negligible, but will be considered later in Sec.~\ref{error}.

In the simulation, we assume that the photon wavepacket have the form
\begin{align}
F(k)=C\times\mathrm{sech}\left[\frac{\omega_k-\omega_{1}}{\sigma_0}\right],
\end{align}
where $C$ is the normalization factor to ensure that
$\int|F(k)|^2\text{d}k=1$, $\omega_{1}$ is the central frequency, and
$\sigma_0$ is the spectral width of the photon wavepacket. To be specific, we take the model system as a doped quantum dot coupled to a rectangle waveguide~\cite{wy}. The two ground states $|1\rangle$ and $|2\rangle$ are represented by the two electron spin states. The excited state $|3\rangle$ is a trion state formed by two electrons in a singlet
state plus a heavy hole. The dipole moment which decides
the coupling strength is assumed
to be 75 Debye and the cut-off frequency of the waveguide is set to
be 1.5 \text{eV}. We consider the sending node with the initial state in $|1\rangle$.

\begin{figure}[tbp]
\center
\includegraphics[width=\linewidth]{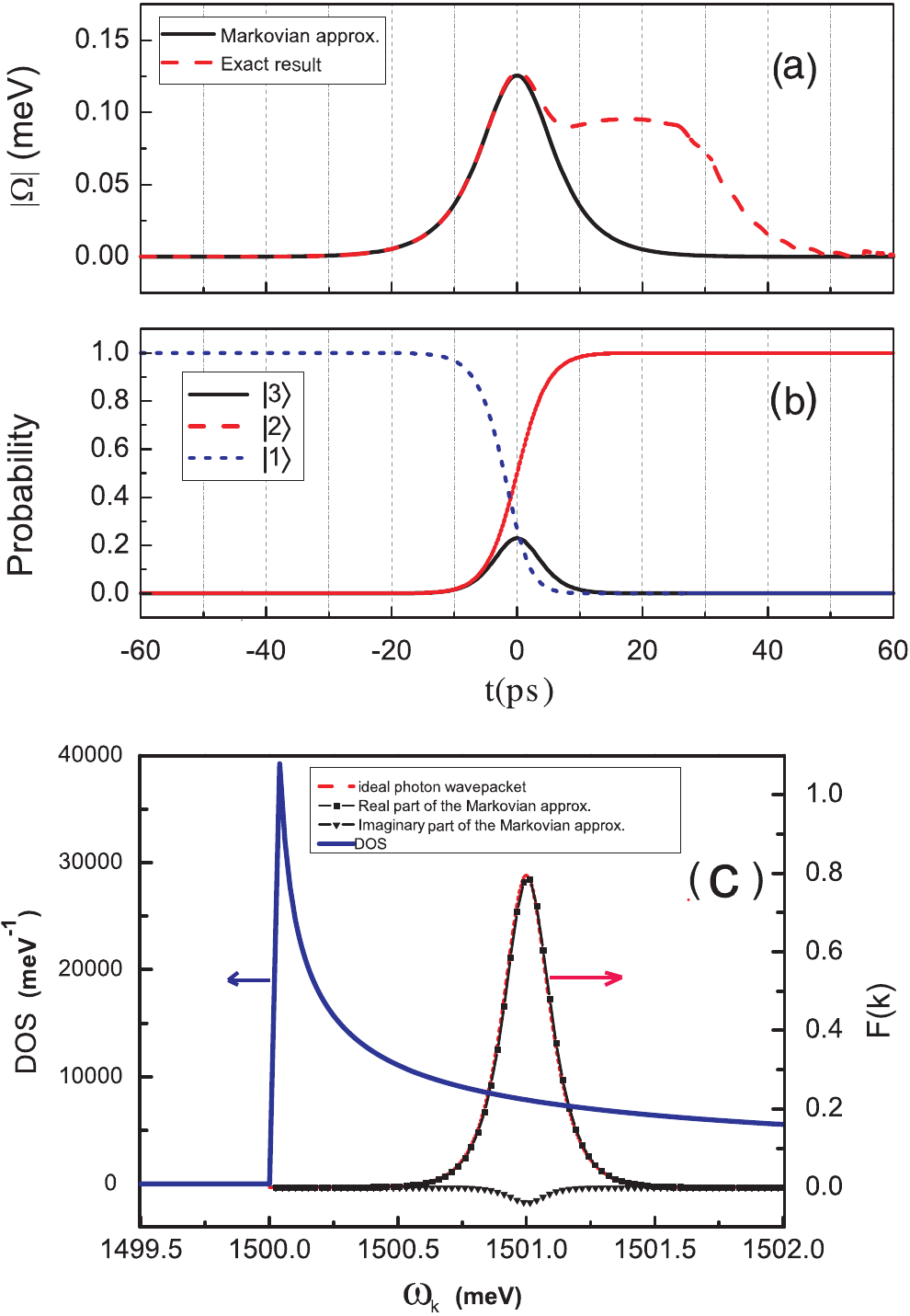}
\caption{(color online) Control of a sending node to emit a photon wavepacket.  The photon wavepacket is 1~meV above the cut-off
frequency of the waveguide and the spectral width is 0.08 meV.
(a) The Rabi frequency of the driving pulse designed with and without Markovian approximation, plot in black solid and red dashed lines, respectively. (b) Populations in the
ground state $|1 \rangle$ (blue dotted line), the ground state $|2
\rangle$ (red dashed line), and the exited state $|3 \rangle$ (black solid line),
 calculated without the Markovian approximation. (c) The outgoing photon wavepacket. The red dashed line is the ideal photon
wavepacket (with zero imaginary part), and the black solid lines with symbols are the outgoing wavepacket obtained with
the Markovian approximation, which has a non-zero imaginary part (black line with triangle symbols). The blue solid line is the DOS in the waveguide.}
\label{fig:omega1}
\end{figure}

Figure~\ref{fig:omega1} compares the control schemes obtained with and without the Markovian approximation. The
central frequency of the outgoing wavepacket is
chosen to be 1~meV above the waveguide cut-off frequency, and $\varepsilon_{32}=1.501~\text{eV}$. The spectral width of the wavepacket is 0.08~meV.
Fig.~\ref{fig:omega1}(a) compares the control pulses obtained with and without the Markovian approximation. At the
beginning
stage of the control process, the two control pulses are almost the
same. In the later stage, the exactly obtained control pulse lasts longer than the Markovian
approximation to complete the full Raman transition. This
can be understood by  memory effects in the photon emission due to the DOS singularity structure in the waveguide continuum. Such memory effects behave as if the emitted photon can be re-absorbed due to the reflection by the waveguide walls. Thus even after the photon emission, a strong laser pulse is still required to avoid the Raman process be reversed.  Fig.~\ref{fig:omega1} (b) shows the state transfer calculated without the Markovian approximation. After about 20~ps, the Raman process for the photon emission is
almost completed and the population of the state $|2 \rangle$ approaches one. Fig.~\ref{fig:omega1} (c)  illustrates the output
 wavepacket calculated with the exact equations of motion [Eq.~(\ref{eqsend})]  but the control pulse designed using
 the Markovian approximation. A small error in photon wavepacket generation is induced as demonstrated by the appearance of a non-zero imaginary part of the wavepacket function.

%\begin{figure}[tbp]
%\center
%\includegraphics[scale=0.80]{1output.eps}
%\caption{The shape of the target pulse, where the target wavepacket
%is 1 meV away above the cut-off frequency with the broadening width
%of 0.08 meV. The red dashed line denotes the ideal $\text{sech}$
%pulse, and the black solid line denotes the outgoing pulse within
%the Markovian approximation method.}
%\label{fig:out1}
%\end{figure}

\begin{figure}[tbp]
\center
\includegraphics[width=\linewidth]{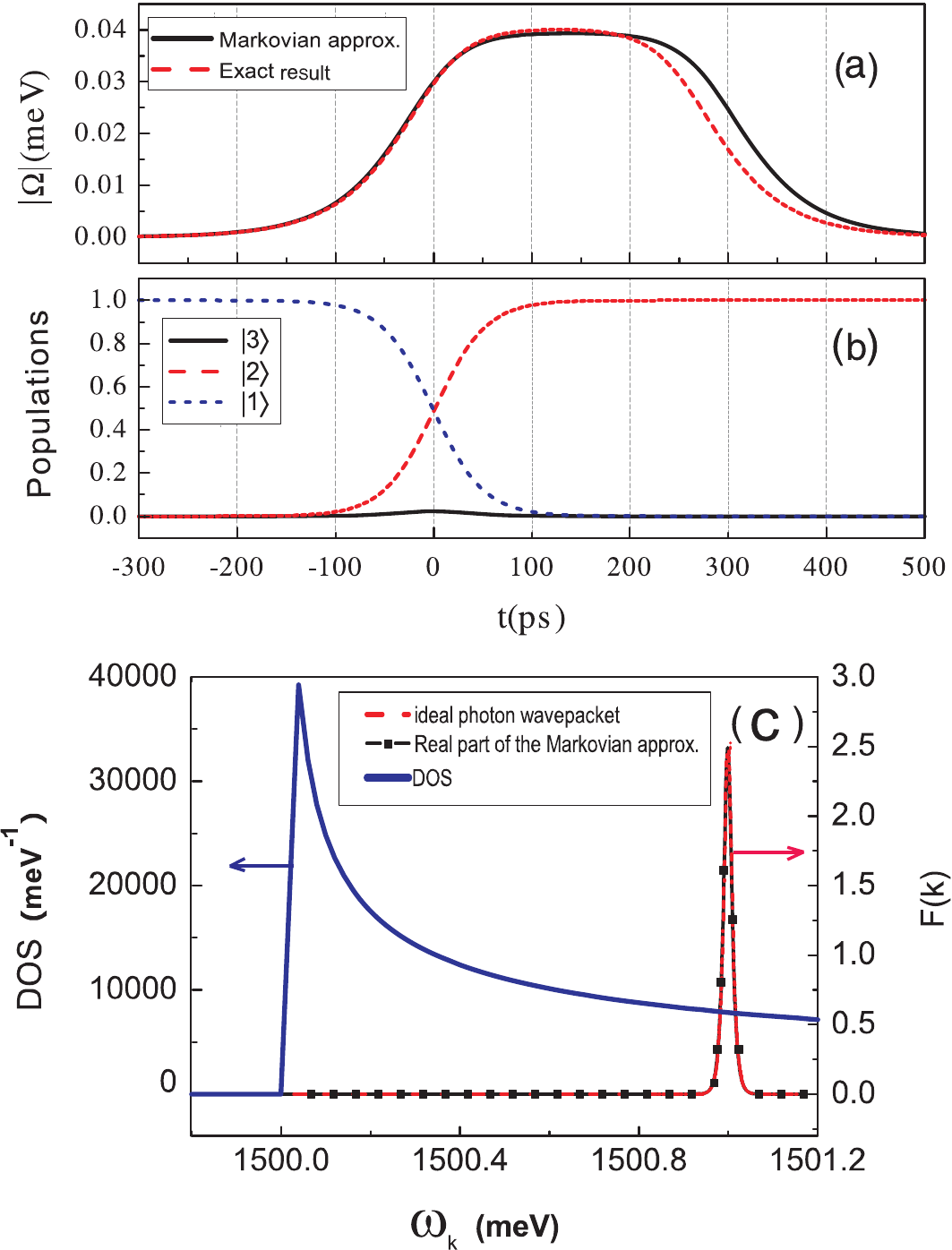}
\caption{(color online) The same as Fig.~\ref{fig:omega1} but the desired photon wavepacket has much smaller spectral width (0.008~meV). Here the imaginary part of the outgoing wavepacket obtained with
the Markovian approximation is equal to zero.}\label{fig:omega2}
\end{figure}

Figure.~\ref{fig:omega2} shows the control under the same conditions as in Fig. ~\ref{fig:omega1} but with the spectral width of the
photon wavepackt to be much smaller ($\sigma_0=0.008$~meV). In this case, the photon emission is much slower than in
Fig.~\ref{fig:omega1}, since the wavepacket duration is inversely proportional to the spectral width. Indeed,
Fig.~\ref{fig:omega2} (b) shows that the
occupation of the exited state during the Raman process is very small, which indicates that
the process is nearly adiabatic. In such a slow process, the Markovian (memory) effect should be less important and the photon re-absorption is negligible. Thus after the photon emission, the control is insensitive to the driving pulse, which just needs to be turned off. As shown in Fig.~\ref{fig:omega2} (a), the Markovian approximation produces almost the same control pulse as the exact solution. The validity of the Markovian approximation in this case can also be understood by the fact that the DOS of the waveguide continuum is nearly flat within the spectral width of the photon wavepacket even though the central frequency is quite close to the cut-off frequency ($\sigma_0\ll \omega_1-\omega_0$). This is seen in Fig.~\ref{fig:omega2} (c) which plots the DOS against the spectrum of the photon wavepackets obtained with and without the Markovian approximation. Fig.~\ref{fig:omega2} (c) also confirms that the photon wavepacket derived with the Markovian approximation is almost the same as the exact result under the slow emission condition.

%==================================================================

\subsection{Effect of photon leakage}
\label{error}
Now we consider the  control errors due to photon leakage into the free space. The leakage reduces the waveguide coupling efficiency. This loss is taken into
account by adding a decay parameter $\gamma\prime$ in the dynamic equations,
\begin{subequations}\label{eqdecay}
\begin{align}
\partial_t{C_1(t)}=&-i\Omega^*(t)C_3(t),\\
\partial_t{C_3(t)}=&-i\Omega(t)C_1(t)-i\int g_kC_2(k)e^{i(\varepsilon_{32}-i\omega_k)t} \text{d}k \nonumber\\
&-\frac{\gamma\prime}{2} C_3(t),\\
\partial_t{C_2(k,t)}=&-ig^*_k C_3(t)e^{i\omega_kt-i\varepsilon_{32}t}.
\end{align}
\end{subequations}
\noindent We define the fidelity of the
sending operation as $|\langle F(k)^{\text{ideal}}|F(k)\rangle|$, where $F(k)^{\text{ideal}}$ is the ideal sech pulse, and $F(k)=C_2(k,t\rightarrow +\infty)$ is calculated by Eq.~(\ref{eqdecay}). The control pulse in Eq.~(\ref{eqdecay}) is obtained by Eq.~(\ref{eqomega}) without the Markovian approximation.

Table.~\ref{tab:fidelity} gives the fidelity for sending a photon wavepacket with various spectral widths and
leakage rates. The parameters are the same as in Fig.~\ref{fig:omega1} and Fig.~\ref{fig:omega2} unless otherwise specified. When the leakage is up to 6 $\%$ of the emission rate to the
waveguide, corresponding to numerical calculation in Ref.~\cite{quan}, the operation fidelity is still above 96$\%$. When the quantum dot is put in the free space, the free-space emission rate is 1$\%$ of the emission rate to the waveguide, and the fidelity is above 99$\%$. The fidelity is determined by the ratio of the decay rate and the leakage rate, insensitive to the spectral width.

\begin{table}[ht]
\caption {Fidelity of the sending operation with finite photon leakage rate ($\gamma'$).  The parameters are the same as in Figs.~2 and 3. For such parameters, the emission rate into the waveguide (estimated with Markovian approximation) is $\gamma$=0.27~meV.} \centering
\begin{tabular}{c c c c c}
  \hline\hline
   spectral width (meV) & 0.08  & 0.08 & 0.008 &0.008 \\
  \hline
  leakage $\gamma^\prime$ & 1 $\% $ $\gamma$   & 6 $\%$ $\gamma$ & 1 $\% $ $ \gamma$ &6 $\%$ $ \gamma$ \\
  \hline
 Fidelity & 0.9916  & 0.9667 & 0.9900 &  0.9606 \\
  \hline
\end{tabular}
\label{tab:fidelity}
\end{table}

%% ==================================================================
%\subsection{Discussion of the state transfer process}
%\label{discussion}
%
%We discuss a general control scheme of a spin-photon quantum
%interface consisting of a three-level system coupled to a continuum
%. The scheme is applicable to solid state systems~\cite{LRB} since the
%three-level systems can couple to the driving field and quantum
%channel via the polarization selection rule~\cite{nanodot}. We extend the exact solution of the control pulse to the non-Markovian regime.
%The  population transfer process clearly shows the pivotal
%time period of the control pulse. We find that Markovian
%approximation only applies for relatively slow pulses, for rapid
%operation near the cut-off frequency of the waveguide,
%non-Markovian dynamics is essential.
%
%The exact solution of the driving pulse encourages the study of the
%fault tolerance of the quantum network.  We also study the effect of
%imperfections such as photon leakage and find high fidelity for the
%operation. With little leakage to the free space, it is possible to
%maintain the fidelity above 99$\%$ during the operation time that
%ranged from 100 ps to 800 ps. The quantum interface with the sending
%part (receiving part) can be considered as a photon source (a
%detector) for arbitrary wavepacket. The applied
%operations include sending, receiving and swapping information
%between spatial distributed nodes.

\section{Strong coupling effect}
\label{photontrap}
When coupled to a continuum, a discrete state becomes a resonance with finite lifetime. If the DOS of the continuum is flat, the discrete state decays exponentially as a typical Markovian process. If the DOS of the continuum varies abruptly, such as the 1D photonic continuum near the cut-off frequency, the resonance of the discrete state does not have a Lorentzian shape. Therefore the non-Markovian effect is important in the dynamics of such systems. If the DOS of the continuum approaches a $\delta$ function, such as in the case of a discrete cavity mode, the coupling leads to the level splitting and the vacuum Rabi oscillation.
A single-mode waveguide has a singularity in the DOS at the lower cut-off frequency, which may replace a discrete state of a cavity. Thus we envisage a scheme of strong coupling between atomic transitions and a waveguide continuum edge. In this regime the decay of an exited atom is
no longer exponential as in the Markovian
approximation.
\subsection{A two-level atom-waveguide model}

The model system is schematically shown in
Fig.~\ref{fig:model} (b). A two-level atom is embedded in the waveguide. $|1\rangle$
and $|0\rangle$ are the exited and ground states of the atom,
respectively. The transition $|1\rangle\leftrightarrow|0\rangle$ is
coupled to the waveguide mode $|k\rangle$ with strength $g$. Once the atom
is in the exited state, it will spontaneously decay by emitting a photon into the waveguide. With the rotating wave
approximation, the Hamiltonian of the system is
\begin{align}
H=\sum_k \hbar \omega_k a^{+}_k a_k +\frac{1}{2}\hbar \omega_{10}
\sigma _z +\hbar \sum_k g(\sigma_+a _k +\sigma_- a^+_k ),\nonumber
\end{align}
where $\omega_{10}$ is the transition frequency of the atom, $\sigma_z=|1\rangle\langle1|-|0\rangle\langle0|$, $\sigma_+=|1\rangle\langle0|$, and $\sigma_-=|0\rangle\langle1|$.  We assume that the coupling strength $g$ is a constant in the following calculation for simplicity.  Here we do not include the free-space emission of the atom and the loss of the waveguide.

\subsection{Theoretical formalism}
\label{Green}
We use the Green's function to study the dynamics of the system. The advanced propagator $G_{+}$ and retarded propagator $G_{-}$ are expressed as~\cite{atomphoton}
\begin{eqnarray}
G_{\pm}(\omega)=\frac{1}{\omega-H\pm i 0^+}.
\end{eqnarray}
%We start from the evolution operator. The evolution operator $\mathscr{U}(t_1,t_2)$ is defined as
%\begin{eqnarray}
%\mathscr{U}(t_1,t_2)=\mathscr{T} \text{exp}[\frac{-i}{\hbar}\int^{t_2}_{t_1}H(t)\text{d} t].
%\end{eqnarray}
%Here $\mathscr{T}$ is the time-ordering operator.
%
%And the transition amplitude between state $|\varphi_1\rangle$ and state $|\varphi_2\rangle$ is determined by
%\begin{eqnarray}
%T_{ij}=\langle \varphi_2 | \mathscr{U}(t_1,t_2) |\varphi_2\rangle .
%\end{eqnarray}
%
%We introduce the Fourier transform of the evolution operator as Green's function, which actually depends only on $t=t_2-t_1$,
%\begin{eqnarray}
%G_{\pm}(\omega)=\text{lim}_{\eta\rightarrow 0_+} G(\omega\pm i \eta)= \frac{1}{i \hbar}\int^{+\infty}_{-\infty} \text{d} t e^{i\omega t/\hbar} \pm \mathscr{U}(t) \theta(\pm t),
%\end{eqnarray}
%where $\theta(t)$ is the Heaviside function.

We assume that initially the atom is in the excited state $|1\rangle$, and the waveguide mode is in vacuum state, i.e., $|\varphi_0\rangle= |1\rangle |\text{vac}\rangle$. The initial state is the eigenstate of the noninteracting
Hamiltonian $H_0=\sum_k \hbar \omega_k a^{+}_k a_k +\frac{1}{2}\hbar \omega_{10}
\sigma _z$. We define the interaction part as $V=H-H_0$.
The matrix element $G(\omega)=\langle\varphi_0 |G|\varphi_0\rangle$ is calculated by summing all perturbative expansion terms of $G(\omega)$ in powers of $V$,
\begin{subequations}
\begin{align}
&G(\omega)=\frac{1}{\omega-\omega_{10}-R(\omega)},
\end{align}
\end{subequations}
where the self energy $R(\omega \pm i 0^+)=\Delta(\omega) \pm i\frac{\Gamma(\omega)}{2}$ with the imaginary part $\Gamma(\omega)$ being the transition rate from the discrete state to the continuum, and the real part $\Delta(\omega)$ the energy shift of the discrete state due to the coupling. The explicit expressions of $\Gamma(\omega)$ and $\Delta(\omega)$ are~\cite{atomphoton}
\begin{subequations}
\begin{align}
&\Gamma(\omega)=2 \pi g^2 \text{DOS}(\omega),\\
&\Delta(\omega)=\frac{1}{2 \pi}\mathscr{P} \int \Gamma(\omega^\prime) \frac{1}{\omega-\omega^\prime} \text{d} \omega^\prime,
\end{align}
\end{subequations}
where $\mathscr{P}$ means the principal part of the integration.

The Green's function gives the evolution operator in the frequency domain,
\begin{eqnarray}\label{greenfunc}
U(\omega)&=&\frac{1}{2\pi i}[G_{-}(\omega)-G_{+}(\omega)]\nonumber\\
&=&\frac{1}{\pi}\frac{{\Gamma(\omega)}/{2}}{[\omega-\omega_{10}-\Delta(\omega)]^2+\left[\Gamma(\omega)/2\right]^2}.
\end{eqnarray}
And the Fourier transform gives the the time-dependence of the probability amplitude in the excited state,
\begin{eqnarray}
U_1(t)=\int \text{d} \omega \mathscr{U}_1(\omega) e^{-i \omega t}.
\label{eq:fourierch3}
\end{eqnarray}
In the weak coupling regime, the denominator in Eq.(~\ref{greenfunc}) is large except in the neighborhood of $\omega_{10}$. We replace $\Delta(\omega)$ and $\Gamma(\omega)$ by $\Delta(\omega_{10})\equiv \Delta$ and $\Gamma(\omega_{10})\equiv \Gamma$. Therefore, the probability amplitude in the excited state is given by the Weisskopf-Wigner approximation,
\begin{equation}
U_1(t)=e^{-i(\omega_{10}+\Delta)t} e^{-\frac{\Gamma}{2} t}.
\end{equation}

\subsection{spontaneous emission in the waveguide}

\begin{figure}
\center
\includegraphics[width=\linewidth]{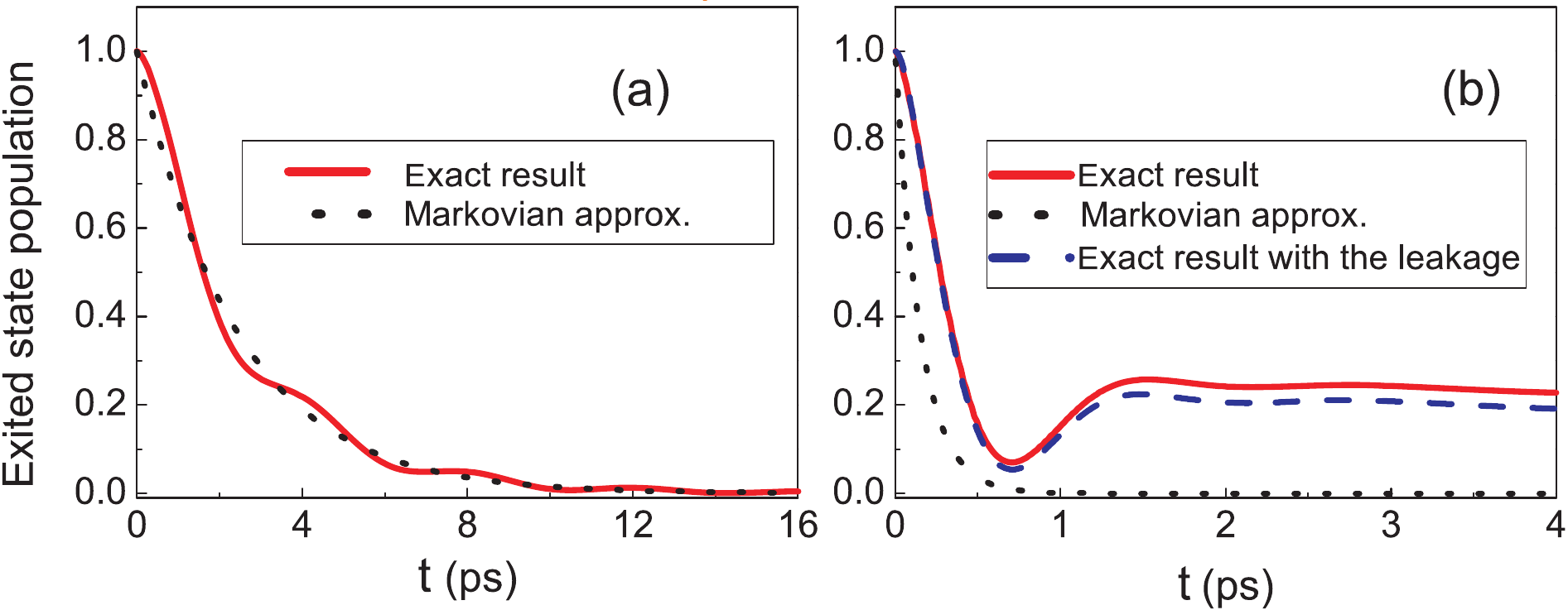}
\caption{ Populations at the excited state of a two-level atom with a dipole moment $p$ of (a) 75 and (b) 300~Debye. The transition frequency is 1~meV above the cut-off frequency (1.5~eV), and the Markovian spontaneous emission rates are 0.27~meV in (a) and 4.37~meV in (b). The red solid lines are the exact results and the black dot lines are the Markovian approximation. The blue dashed line in (b) is the exact result with the photon leakage rate of 0.033~meV. }\label{fig:excited}
\end{figure}

Figure~\ref{fig:excited} compares the spontaneous emission obtained by Markovian approximation with the exact results, for various atom dipole moments. In the strong coupling regime, the vacuum Rabi oscillation is observed [Fig.~\ref{fig:excited} (b)], which we ascribe to the splitting between a discrete bound state and a DOS singularity of the photon continuum. At the long-time limit, the oscillation is damped but some residual population at the exited state of the atom is still left. Such incomplete emission evidences the formation of a bound polariton state. The residual atomic population increases as the coupling strength increases, on the account of the enhancement of the atomic component of the polariton state.

\begin{figure}
\center
\includegraphics[width=\linewidth]{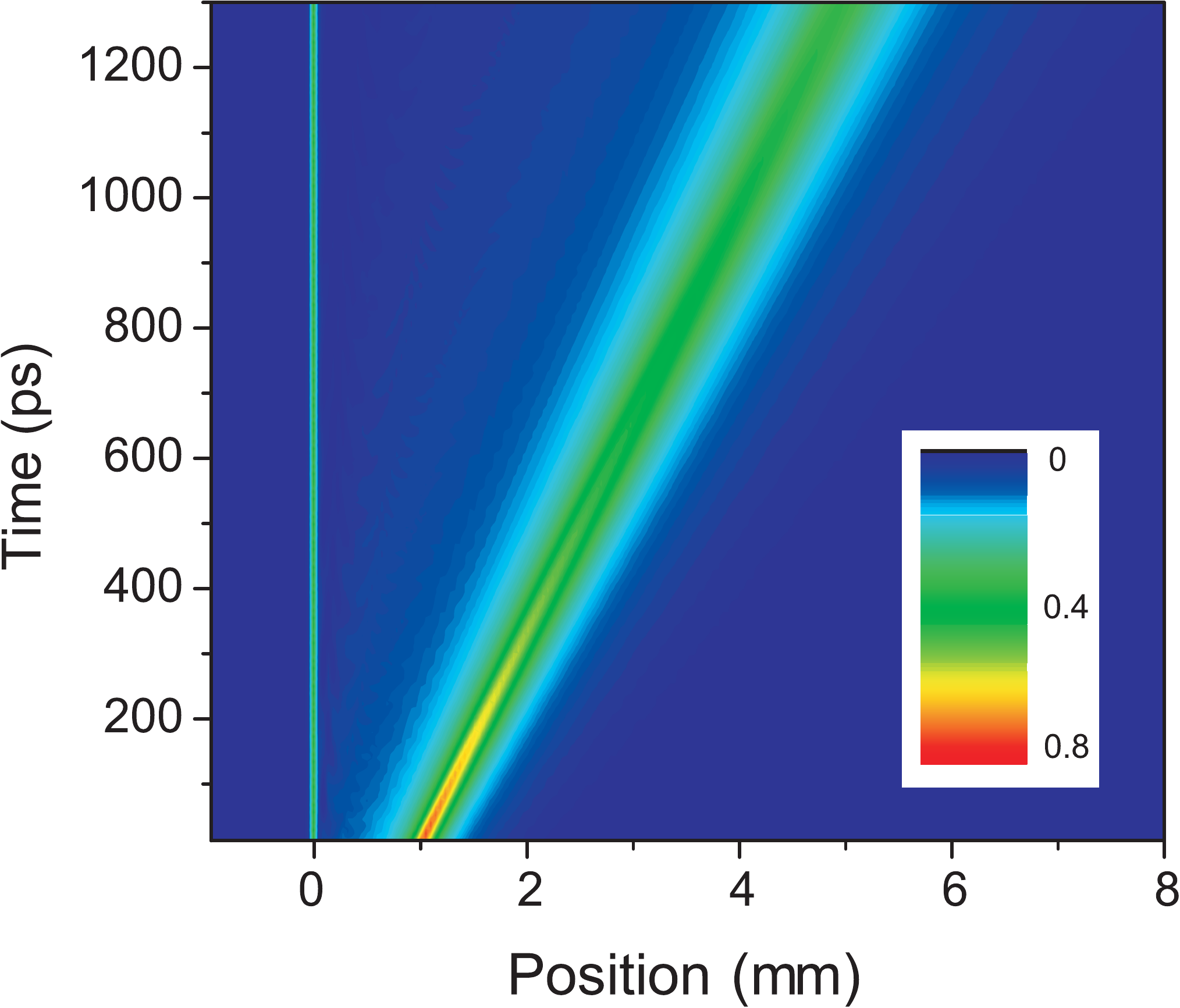}
\caption{(color online) The photon wavepacket amplitude in the waveguide at various times after the trigger of the emission at $t=0$. The atom is placed at $x$ = 0. }\label{fig:pwavepacket}
\end{figure}

The formation of a bound polariton state leads to the photon localization. To show this effect, we calculate the real-time evolution of the photon wavepacket during the emission and propagation (Fig.~\ref{fig:pwavepacket}). We expand the state $|\varphi(t)\rangle$ of the two-level system into the following superposition,
\begin{eqnarray}
|\varphi(t)\rangle=C_1(t)|1\rangle|\text{vac}\rangle+\int \text{d}k C_0(k,t) |0\rangle|k\rangle.
\end{eqnarray}
And the photon wavepacket $f(x,t)$ is calculated by the Fourier transform
\begin{eqnarray}
f(x,t)=\int \text{d}k C_0(k,T_0)e^{-i \omega_k(t-T_0)}e^ {i k x},
\end{eqnarray}
where $T_0$ is the final time of the emission taken as 10~ps.  As shown in Fig.~\ref{fig:pwavepacket} the photon wavepacket propagates slowly after emission (at a speed only 1$\%$ that of light in vacuum), and  some portion of the wavepacket is localized at the atom position, which is related to the photon component in the bound polariton state.

When the realistic leakage emission into free space is considered, the bound polariton state has a finite lifetime and the residual atomic population will eventually decay to zero after the photon emission into the waveguide. The numerical calculation shows that when the ratio of free-space emission to the waveguide emission is less than 1$\%$, the free-space leakage makes no significant changes to the strong coupling effects as shown in Fig.\ref{fig:excited} (b).

%\subsection{Conclusion about the photon trapping in the waveguide}
%We explored the photon emission of an atom in a waveguide and in particular studied the effect of the DOS singularity of the photon continuum near the cut-off frequency. We found the photon localization effect owing to the formation of a trapped polariton. This effect realizes strong atom-photon coupling without a cavity, with the DOS singularity in a waveguide mimicking a discrete state in a cavity.

\section{Conclusion}
\label{conclusion}

In summary, we have studied a general control scheme of a spin-photon quantum
interface consisting of a three-level system coupled to a 1D photonic continuum
and extended the exact solution of the control scheme~\cite{wy} to the non-Markovian regime. We found that the Markovian approximation only applies to relatively slow operation. For rapid
operation near the cut-off frequency of the waveguide,
the non-Markovian dynamics is essential. We have also studied the effect of
imperfections such as photon leakage and found high fidelity for the
operation in presence of realistic free-space emission. With realistic leakage to the free space of $6\%$, it is possible to
maintain the fidelity above 96$\%$ for an operation time
ranging from 100 ps to 800 ps.

In addition, for a coupled atom-waveguide system, we have studied the photon emission of two-level atoms and in particular the effect of the DOS singularity at the photon continuum edge. We show the vacuum Rabi oscillation and the formation of a bound polariton state, which reflect the strong coupling between a discrete atomic transition and a photonic continuum edge. We found the photon localization effect due to the formation of a trapped polariton. This effect realizes strong atom-photon coupling without a cavity, with the DOS singularity in a waveguide mimicking a discrete state in a cavity.

\section*{ACKNOWLEDGEMENT}
This work was supported by Hong Kong RGC CUHK402209 and Focused Investments Scheme of The Chinese University of Hong Kong.
%==================================================================

%\bibliography{references}

\end{document}